\documentstyle[preprint,eqsecnum,aps,amsfonts,amssymb,psfig,epsfig,prb]{revtex} 

\begin{document}

%% \date{February 26, 2002}

\def\reff#1{(\ref{#1})}
\newcommand{\be}{\begin{equation}}
\newcommand{\ee}{\end{equation}}
\newcommand{\<}{\langle}
\renewcommand{\>}{\rangle}

%%%  \ltapprox and \gtapprox produce > and < signs with twiddle underneath
\def\spose#1{\hbox to 0pt{#1\hss}}
\def\ltapprox{\mathrel{\spose{\lower 3pt\hbox{$\mathchar"218$}}
 \raise 2.0pt\hbox{$\mathchar"13C$}}}
\def\gtapprox{\mathrel{\spose{\lower 3pt\hbox{$\mathchar"218$}}
 \raise 2.0pt\hbox{$\mathchar"13E$}}}

\def\bsigma{\mbox{\protect\boldmath $\sigma$}}
\def\bpi{\mbox{\protect\boldmath $\pi$}}
\def\smfrac#1#2{{\textstyle\frac{#1}{#2}}}
\def\smhalf{ {\smfrac{1}{2}} }

\newcommand{\re}{\mathop{\rm Re}\nolimits}
\newcommand{\im}{\mathop{\rm Im}\nolimits}
\newcommand{\tr}{\mathop{\rm tr}\nolimits}
\newcommand{\fr}{\frac}
\newcommand{\diti}{\frac{\mathrm{d}^2t}{(2 \pi)^2}}
\newcommand{\bz}{{\mathbf 0}}

\def\Z{{\mathbb Z}}
\def\R{{\mathbb R}}
\def\C{{\mathbb C}}

\title{Virial coefficients and osmotic pressure in polymer
solutions \\ in good-solvent conditions}

\author{ Sergio Caracciolo, Bortolo Matteo Mognetti}
\address{Dipartimento di Fisica and INFN -- Sezione di Milano I
  Universit\`a degli Studi di Milano \\
  Via Celoria 16, I-20133 Milano, Italy \\  
  e-mail: {\tt Sergio.Caracciolo@sns.it},
          {\tt Bortolo.Mognetti@mi.infn.it}}
\author{ Andrea Pelissetto }
\address{
  Dipartimento di Fisica and INFN -- Sezione di Roma I  \\
  Universit\`a degli Studi di Roma ``La Sapienza'' \\
  P.le A. Moro 2, I-00185 Roma, Italy \\
  e-mail: {\rm Andrea.Pelissetto@roma1.infn.it }}

\maketitle
\thispagestyle{empty}   % Suppress page number on front page.

\begin{abstract}

We determine the second, third, and fourth virial coefficients appearing
in the density expansion of the 
osmotic pressure $\Pi$ of a monodisperse polymer solution in 
good-solvent conditions. Using the expected large-concentration behavior, 
we extrapolate the low-density expansion outside the 
dilute regime, obtaining the osmotic pressure for any concentration in
the semidilute region. Comparison with field-theoretical predictions 
and experimental data shows that the obtained expression is quite accurate.
The error is approximately 1-2\% below the overlap concentration
and rises at most to 5-10\% in the limit of very large polymer 
concentrations.

\bigskip 


% 61.25.Hq  Macromolecular and polymer solutions; polymer melts; swelling 
% 82.35.Lr  Physical properties of polymers 

\end{abstract}

\clearpage

\section{Introduction}

For sufficiently high molecular weights, dilute and semidilute 
polymer solutions under good-solvent conditions exhibit a universal 
scaling behavior.\cite{deGennes-72,deGennes-79,Freed-87,Schaefer-99}
For instance, the radius of gyration $R_g$, which gives the average size
of the polymer, scales as $N^\nu$, where $N$ is the degree of 
polymerization and $\nu$ a universal exponent, $\nu \approx 0.5876$
(Ref.~\CITE{bestnu}). The osmotic pressure $\Pi$ is one of the 
most easily accessible quantities in polymer physics. 
When $N$ is large, it obeys a general scaling 
law\cite{deGennes-79,Freed-87,Schaefer-99} (here and in the following
we only consider monodisperse solutions)
\begin{equation}
Z \equiv   {M \Pi \over R T \rho} = {\Pi\over k_B T c} = f (c R_g^3),
\label{scalingZ}
\end{equation}
where $c$ is the polymer number density, $\rho$ the ponderal concentration,
$M$ the molar mass of the polymer, and $T$ the absolute 
temperature. The function $f(x)$ is universal, so that the 
determination of $Z$ in a specific model allows one to predict
$\Pi$ for any polymer solution. In the dilute limit the 
compressibility factor $Z$ can be expanded in powers of $c$ 
obtaining\cite{footexpt}
\begin{equation}
Z = 1 + \sum_{n=1} B_{n+1} c^n,
\label{expanZ1}
\end{equation}
where the coefficients $B_n$ are knows as virial coefficients. Knowledge of 
$B_n$ allows one to compute $\Pi$ in the dilute regime in which 
$c R^3_g \ll 1$. The coefficients $B_n$ depend on the polymer solution.
Eq.~(\ref{expanZ1}) can be rewritten as 
\begin{equation}
Z = 1 + \sum_{n=1} A_{n+1} (c R_g^3)^n \qquad\qquad
A_{n+1} \equiv  B_{n+1} R_g^{-3n},
\label{expanZ}
\end{equation}
where $R_g$ is the zero-density radius of gyration of the polymer.
The general scaling law (\ref{scalingZ}) implies that for 
$N\to \infty$ the coefficients $A_{n+1}$ approach 
universal constants $A_{n+1}^*$ that are independent of chemical 
details.
The value of $A_2^*$ has been the object of many theoretical studies 
(sometimes, the interpenetration radius $\Psi \equiv 2 (4\pi)^{-3/2} A_2$ is 
quoted instead of $A_2$). The most precise estimates have been 
obtained by means of Monte Carlo (MC) simulations:
$A^*_2 = 5.494\pm 0.005$ (Ref.~\CITE{PH-05}),
$A^*_2 = 5.504\pm 0.007$ (Ref.~\CITE{LMS-95}),
$A^*_2 = 5.490\pm 0.027$ (Ref.~\CITE{Nickel-91}).
Eq.~(\ref{scalingZ}) 
is  only valid for high molecular weights. 
Since in many cases $N$ is not very large, it is important to 
consider also the leading correction to this expression. As predicted 
by the renormalization group and extensively verified numerically, 
for $N$ large but finite we have 
\begin{equation}
Z \approx  f(c R^3_g) + {a_2 \over N^\Delta} g(c R^3_g), \qquad\qquad
A_{n+1} \equiv  {B_{n+1}\over R^{3n}_g} \approx 
     A_{n+1}^* + {a_2 b_{n+1} \over N^\Delta},
\label{corrections}
\end{equation}
where $b_2 = 1$, $\Delta$ is a universal exponent whose best estimate 
is\cite{BN-97} $\Delta = 0.515\pm 0.007^{+0.010}_{-0.000}$,
and, of course, $g(x) = 1 + \sum b_{n+1} x^n$.
The function $g(x)$ as well as the constants $b_n$ are universal.
All chemical details as well as polymer properties---for instance,
the temperature---are encoded in a single constant $a_2$ that 
varies from one polymer solution to the other. 
In this paper we extend the previous 
calculations to the third and fourth virial coefficient, computing 
$A_3^*$, $A_4^*$, and $b_3$.
For this purpose we perform an extensive MC simulation
of the lattice Domb-Joyce model,\cite{DJ-72} considering walks 
of length varying between 100 and 8000 and three different 
penalties for the self-intersections. 

Knowledge of the first virial coefficients and of the leading 
scaling corrections
allows us to obtain a precise prediction for the osmotic pressure in the 
dilute regime (the expression is apparently accurate up 
to $B_2 c \approx 1$), 
even for relatively small values of the degree of polymerization.
Finite-length effects are taken into account by properly tuning a single 
nonuniversal parameter.  Once the virial expansion is 
known, we can try to resum it to obtain an 
interpolation formula that is valid in the semidilute regime. 
We will show that a simple expression that takes into account 
the large-density behavior of $Z$ provides a good approximation to
$Z$, even outside the dilute regime. 

The paper is organized as follows. In Sec.~\ref{sec2} we derive the 
virial expansion for a polymer solution. In Sec.~\ref{sec3} we define the 
model, while in Sec.~\ref{sec4} we compute the universal constants
defined above that are associated with the virial coefficients.  
In Sec.~\ref{sec5} we present our conclusions and, in particular, 
give an interpolation formula for $Z$ that is also valid in the 
semidilute regime. Some technical details are presented in the Appendix.

\section{Virial expansion} \label{sec2}

We wish now to derive the virial expansion for a polymer solution.
Such an expansion is easily derived in the grand-canonical 
ensemble.\cite{HMD-76} For this purpose, we first define 
the configurational partition function $Q_L$ of 
$L$ polymers in a volume $V$:
\begin{eqnarray}
Q_L&=& \int d\mu\, \exp\left(-\beta\sum_{i<j} V_{ij}^{\rm inter}\right)
\\
d\mu &=& \left[\prod_{i\alpha} d^3{\mathbf r}_{\alpha i}\right] 
   \exp\left(-\beta\sum_{i} V_{i}^{\rm intra}\right),
\end{eqnarray}
where $V_{ij}^{\rm inter}$ is the sum of all terms of the Hamiltonian
that correspond to interactions of monomers belonging to two
different polymers $i$ and $j$, 
$V_{i}^{\rm intra}$ is the contribution due to interactions of monomers
belonging to the same polymer $i$, and ${\mathbf r}_{\alpha i}$ is 
the position of monomer $\alpha$ belonging to polymer $i$. 
Then, we set 
\be
\Xi = \sum_L {z^L\over L!} Q_L.
\ee
The virial expansion is obtained by performing an expansion in powers of 
$z$. We introduce the Mayer function
\begin{equation}
f_{ij} \equiv  \exp(- \beta V_{ij}^{\rm inter}) - 1
\end{equation}
and define the following integrals:
\begin{eqnarray}
I_2 &\equiv & \int d^3 {\mathbf r}_{12} \, 
       \langle f_{12} \rangle_{{\mathbf 0},{\mathbf r}_{12}} 
\\
I_3 &\equiv & 
    \int d^3{\mathbf r}_{12} d^3{\mathbf r}_{13}
     \left\langle f_{12}f_{13}f_{23}
     \right\rangle_{{\mathbf 0},{\mathbf r}_{12},{\mathbf r}_{13}} 
\\
I_4 &\equiv & 
     \int d^3{\mathbf r}_{12} d^3{\mathbf r}_{13} d^3{\mathbf r}_{14}
    \left\langle f_{12}f_{23}f_{34}f_{14}(3+6 f_{13}+ f_{13}f_{24})
\right\rangle_{{\mathbf 0},{\mathbf r}_{12},{\mathbf r}_{13},{\mathbf r}_{14}}
\\
T_1 &\equiv & \int d^3{\mathbf r}_{12} d^3{\mathbf r}_{13}
      \left\langle f_{12} f_{13} 
       \right\rangle_{{\mathbf 0},{\mathbf r}_{12},{\mathbf r}_{13}}
   -\left[ \int d^3{\mathbf r}_{12}
    \left\langle f_{12} \right\rangle_{{\mathbf 0},{\mathbf r}_{12}}\right]^2
\\
T_2 &\equiv & 
   \int d^3{\mathbf r}_{12} d^3{\mathbf r}_{13} d^3 {\mathbf r}_{14}  
    \left\langle f_{12} f_{13} f_{14} \right
    \rangle_{{\mathbf 0},{\mathbf r}_{12}, {\mathbf r}_{13},{\mathbf r}_{14}}
   -\left[ \int d^3{\mathbf r}_{12}\left\langle f_{12} 
           \right\rangle_{{\mathbf 0},{\mathbf r}_{12}}\right]^3
\\
T_3 &\equiv &\int d^3{\mathbf r}_{12} d^3{\mathbf r}_{13} d^3 {\mathbf r}_{14}  
   \left\langle f_{12} f_{23} f_{34} \right
   \rangle_{{\mathbf 0},{\mathbf r}_{12}, {\mathbf r}_{13},{\mathbf r}_{14}}
  -\left[ \int d^3{\mathbf r}_{12}\left\langle f_{12} 
          \right\rangle_{{\mathbf 0},{\mathbf r}_{12}}\right]^3
\\
T_4 &\equiv &\int d^3{\mathbf r}_{12} d^3{\mathbf r}_{13} d^3 {\mathbf r}_{14}  
\left\langle f_{12} f_{23} f_{13 }f_{34} \right
\rangle_{{\mathbf 0},{\mathbf r}_{12}, {\mathbf r}_{13},{\mathbf r}_{14}}
\nonumber\\
&&-\left[ \int d^3{\mathbf r}_{12}\left\langle f_{12} 
   \right\rangle_{{\mathbf 0},{\mathbf r}_{12}}\right]
   \left[  \int d^3{\mathbf r}_{12}d^3{\mathbf r}_{13}
   \left\langle f_{12} f_{13} f_{23} 
         \right\rangle_{{\mathbf 0},{\mathbf r}_{12},{\mathbf r}_{13}}\right].
\end{eqnarray}
Here $\langle \cdot \rangle_{{\mathbf 0},{\mathbf r}}$ indicates an average 
over two independent polymers such that the first one starts at the 
origin and the second starts in ${\mathbf r}$. Analogously 
$\langle \cdot \rangle_{{\mathbf 0},{\mathbf r}_2,{\mathbf r}_3}$ and 
$\langle \cdot \rangle_{{\mathbf 0},{\mathbf r}_2,{\mathbf r}_3,{\mathbf r}_4}$
refer to averages over three and four polymers respectively, 
the first one starting in the origin, the second in 
${\mathbf r}_2$, etc. 

Then, a simple calculation gives 
\begin{eqnarray}
{\beta \Pi} = {1\over V} \ln \Xi &=& 
   z + {z^2\over2} I_2 + {z^3\over 6} (I_3 + 3 T_1 + 3 I_2^2) 
\nonumber \\
   && + {z^4\over24} (I_4 + 4 T_2 + 12 T_3 + 12 T_4 + 16 I_2^3 + 
       12 I_2 I_3) + O(z^5).
\label{Pi-z}
\end{eqnarray}
The density is obtained by using 
\be
c = z {\partial \beta\Pi\over \partial z}.
\ee
The previous equation can be inverted to obtain $z$ in powers of $c$.
Substituting in Eq.~(\ref{Pi-z}), we obtain expansion \reff{expanZ1}
with 
\begin{eqnarray}
B_2 &=& - {1\over 2} I_2  ,
\label{B2def} \\
B_3 &=& - {1\over 3} I_3  - T_1 ,
\label{B3def} \\
B_4 &=& - {1\over 8} I_4  -
          {1\over 2} T_2 - {3\over 2} T_3 - {3\over 2} T_4 + 
          {9\over 2} I_2 T_1 .
  \label{B4def}
\end{eqnarray}
Note that there are additional contributions to $B_3$ and $B_4$ which 
are missing in simple fluids.\cite{HMD-76} Indeed, for a monoatomic fluid
$T_n = 0$. These terms are instead present
in the polymer virial expansion. 

In the following we shall consider a lattice model for polymers. 
In this case, the previous expressions must be trivially modified, 
replacing each integral with the corresponding sum over all 
lattice points. 

\section{Model and observables} \label{sec3}

Since we are interested in computing the universal quantities 
$A_n^*$ and $b_n$, we can use any model that captures the basic 
polymer properties. For computational convenience we consider 
a lattice model.  A polymer of length
$N$ is modelled by a random walk 
$\{{\mathbf r}_0,{\mathbf r}_1,\ldots,{\mathbf r}_N\}$ with 
$|{\mathbf r}_\alpha-{\mathbf r}_{\alpha+1}|=1$ on a cubic lattice. 
To each walk we associate a Boltzmann factor
\begin{equation}
e^{-\beta H} = e^{-w \sigma},\qquad\qquad
\sigma = \sum_{0\le \alpha < \beta \le N} 
   \delta_{{\mathbf r}_\alpha,{\mathbf r}_\beta},
\end{equation}
with $w \ge 0$. The factor $\sigma$ counts how many 
self-intersections are present in the walk. This model is similar
to the standard self-avoiding walk (SAW) model in which polymers are 
modelled by random walks in which self-intersections are forbidden. 
The SAW model is obtained for $w = +\infty$. For finite positive $w$ 
self-intersections are possible although energetically penalized. 
For any positive $w$, this model---hereafter we will refer to it
as Domb-Joyce (DJ) model---has the same scaling limit of the 
SAW model\cite{DJ-72} and thus allows us to compute the 
universal scaling functions that are relevant for polymer solutions.
The DJ model has been extensively studied numerically in Ref.~\CITE{BN-97}.
There, it was also shown that there is a particular value of $w$,
$e^{-w^*} \approx 0.603$ (i.e.,
$w^* \approx 0.505838$), for which corrections to scaling 
with exponent $\Delta$ vanish: the nonuniversal constant 
$a_2$ is zero for $w=w^*$. Thus, simulations
at $w = w^*$ are particularly convenient since the scaling limit
can be observed for smaller values of $N$.

In the simulations we measure the virial coefficients
using Eqs.~(\ref{B2def}), (\ref{B3def}),
and (\ref{B4def}). In this model  the Mayer function is simply
\begin{eqnarray}
&& f_{ij} = e^{-\beta V_{ij}^{\rm inter}} - 1 = e^{-w \sigma_{ij}} - 1
\\
&& \sigma_{ij} = \sum_{\alpha\beta} 
   \delta_{{\mathbf r}_{\alpha i},{\mathbf r}_{\beta j}}.
\end{eqnarray}
Here ${\mathbf r}_{\alpha i}$ is the position of monomer $\alpha$ of polymer
$i$. 
The DJ model can be efficiently simulated by using the pivot 
algorithm.\cite{Lal,MacDonald,Madras-Sokal,Sokal-95b} 
For the SAW an efficient implementation is discussed in 
Ref.~\CITE{Kennedy-02}. The extension to the DJ model is straightforward,
the changes in energy being taken into account by means of 
a Metropolis test. Such a step should be included carefully
in order not to loose the good scaling behavior of the CPU time 
per attempted move. We use here the implementation discussed in 
Ref.~\CITE{CPP-94}.
The virial coefficients have been 
computed by using a simple generalization of the hit-or-miss 
algorithm discussed in Ref.~\CITE{LMS-95}. Some details 
are reported in the Appendix. 

\section{Monte Carlo determination of the virial coefficients} \label{sec4}

We perform three sets of simulations at $w = 0.375$,
0.505838, 0.775, using walks with $100 \le N \le 8000$. 
Results are reported in Tables~\ref{tableA2},
\ref{tableA3}, and \ref{tableA4}. 
As expected, the results for $w = 0.505838$ are the least dependent on 
$N$, confirming that for this value of $w$ scaling corrections are 
very small. On the other hand, for $w = 0.375$ and $w = 0.775$ scaling 
corrections are sizable.

The numerical data are analyzed as discussed in Ref.~\CITE{PH-05}. 
We assume that $A_n$ has an expansion of the form 
\begin{equation}
A_n (N,w) \approx A_n^* + {a_n(w)\over N^\Delta} + 
                    {c_n(w)\over N^{\Delta_{2,\rm eff}}}\; .
\label{Anfit}
\end{equation}
For $\Delta$ we use the best available estimate: 
$\Delta = 0.515\pm 0.007^{+0.010}_{-0.000}$ (Ref.~\CITE{BN-97}).
The term $1/N^{\Delta_{2,\rm eff}}$ should take into account analytic
corrections behaving as $1/N$, and nonanalytic ones of the form
$N^{-2\Delta}$, $N^{-\Delta_2}$ ($\Delta_2$ is the next-to-leading
correction-to-scaling exponent). As discussed in Ref.~\CITE{PH-05},
one can lump all these terms into a single one with exponent 
$\Delta_{2,\rm eff} = 1.0\pm 0.1$. In order to estimate $A_2$
we also use the results that appear in Table 1 of Ref.~\CITE{PH-05}
that refer to MC simulations of interacting SAWs.\cite{footA3} 
A combined fit of
the results for $A_2$ (with $1 + 5\times 2 = 11$ free parameters) gives
\begin{eqnarray}
A_2^* = 5.4986 \pm 0.0010 \pm 0.0002  && 
           \qquad\qquad N_{\rm min} = 250, \nonumber \\
A_2^* = 5.4997 \pm 0.0017 \pm 0.0002  &&  
           \qquad\qquad N_{\rm min} = 500, \nonumber \\
A_2^* = 5.5013 \pm 0.0030 \pm 0.0003  &&  
           \qquad\qquad N_{\rm min} = 1000 \; .
\end{eqnarray}
In each fit we have considered only the data with $N\ge N_{\rm min}$.
We do not show results for $N_{\rm min} = 100$ since in this case the fit has a 
somewhat large $\chi^2$. We report two error bars. The first one 
is the statistical error while the second gives the variation of the 
estimate as $\Delta$ and $\Delta_{2,\rm eff}$ vary within one error bar.
The results show a small upward trend which is in any case of the order of the 
statistical errors. 
As our final estimate we quote 
\begin{equation}
A_2^* =  5.500\pm 0.003.
\label{A2star}
\end{equation}
Note that in the polymer literature one often considers the interpenetration
ratio $\Psi\equiv 2 (4\pi)^{-3/2} A_2$ instead of $A_2$. We have
\be 
\Psi^* = 2 (4\pi)^{-3/2} A_2^* = 0.24693\pm  0.00013.
\ee
The same analysis---but in this case we only rely on the DJ results---can be 
repeated for $A_3$. Estimates of $A_3$ are reported in Table \ref{tableA3}.
The contribution proportional to $I_3$ appearing in Eq.~(\ref{B3def})---the 
only one present in simple fluids--- accounts for 
most of the result since the contribution proportional to 
$T_1$ is small, $T_1 R_g^{-6} \approx 0.84$ in the scaling limit. Still 
in a high-precision calculation, $T_1$ cannot be neglected, giving 
a 9\% correction. A fit of the results to Eq.~(\ref{Anfit}) gives
\begin{eqnarray}
A_3^* = 9.788 \pm 0.005 \pm 0.002 &&  
    \qquad\qquad N_{\rm min} = 100, \nonumber \\
A_3^* = 9.786 \pm 0.008 \pm 0.001 &&  
    \qquad\qquad N_{\rm min} = 250, \nonumber \\
A_3^* = 9.798 \pm 0.013 \pm 0.001 &&  
    \qquad\qquad N_{\rm min} = 500, \nonumber \\
A_3^* = 9.813 \pm 0.015 \pm 0.001 &&  
    \qquad\qquad N_{\rm min} = 1000 \; ,
\end{eqnarray}
where, as before, the first error is the statistical one while the 
second is related to the error on $\Delta$ and $\Delta_{\rm eff}$.
As final estimate we quote 
\begin{equation}
A_3^* = 9.80 \pm 0.02\; .
\label{A3star}
\end{equation}
In the theoretical literature, several estimates have been reported for 
the large-$N$ value of $A_{3,22} \equiv B_3/B_2^2$ 
(this quantity is often called $g$), 
which is universal and 
independent of the radius of gyration. 
Using Eqs.~(\ref{A2star}) and (\ref{A3star}) we obtain
\begin{equation}
A_{3,22}^* = A_3^*/(A_2^*)^2 = 0.3240 \pm 0.0007.
\end{equation}
Finally, we consider $A_4$. In this case statistical errors are quite large.
This is due to significant cancellations among the different terms appearing 
in Eq.~(\ref{B4def}). Moreover, while in $A_3$ the term $T_1$ was providing
only a small correction, here inclusion of the terms proportional 
to $T_i$ is crucial to obtain the correct result. They are not small:
in the scaling limit we have
$T_2 R_g^{-9} \approx 28$,  $T_3 R_g^{-9} \approx 18$,
$T_4 R_g^{-9} \approx 1.5$, $T_1 I_2 R_g^{-9} \approx -9$.
Fits to Eq.~(\ref{Anfit}) give
\begin{eqnarray}
A_4^* =  -9.2 \pm 0.5   &&  \qquad\qquad N_{\rm min} = 100, \nonumber \\
A_4^* =  -9.3 \pm 0.7   &&  \qquad\qquad N_{\rm min} = 250, \nonumber \\
A_4^* =  -9.6 \pm 1.3   &&  \qquad\qquad N_{\rm min} = 500.
\end{eqnarray}
The systematic error is negligible in this case. 
In order to improve the result we have repeated the analysis taking into 
account that $a_4(w) = b_4 a_2(w)$, with $b_4$ independent of $w$.
If we analyze together the data for $A_2$ and $A_4$ taking as free parameters
$A_2^*$, $A_4^*$, $b_4$, $a_2(w_i)$, $c_2(w_i)$, and $c_4(w_i)$, the nonlinear 
fit gives:
\begin{eqnarray}
A_4^* =  -8.89 \pm 0.29   &&  \qquad\qquad N_{\rm min} = 250 , \nonumber \\
A_4^* =  -9.00 \pm 0.36   &&  \qquad\qquad N_{\rm min} = 500 .
\end{eqnarray}
Correspondingly, we obtain $b_4 = -9\pm 7$ and $b_4 = -2\pm 11$. 
Comparing all results we obtain
\begin{equation}
A_4^* = -9.0 \pm 0.5.
\end{equation}
Our result for $A_2^*$ is in good agreement with previous MC estimates:
$A^*_2 = 5.494\pm 0.005$ (Ref.~\CITE{PH-05}),
$A^*_2 = 5.504\pm 0.007$ (Ref.~\CITE{LMS-95}),
$A^*_2 = 5.490\pm 0.027$ (Ref.~\CITE{Nickel-91}).
Direct MC calculations\cite{footA3} of $A_3^*$ 
provided only the order of magnitude since they 
did not consider the contribution proportional to 
$T_1$. Ref.~\CITE{SOKK-00} quotes an estimate of 
$A_{3,22}$, $A_{3,22} \approx 0.40$-$0.45$. This value is somewhat higher 
than  what we obtain here, but it must be noted that very short walks 
were considered ($N\le 50$). 
Thus, those results are probably affected by strong scaling 
corrections. Overall, our 
estimate of $A_{3,22}$ is in agreement with the field-theory estimates
\cite{dCN-82,DF-85,Schaefer-99}
that vary between 0.28 and 0.43.
A field-theory estimate of $A_4^*$ can be obtained from the expressions
\cite{footA4} presented in 
Sec.~17.3.2 of Ref.~\CITE{Schaefer-99}. The result, 
$A_4^* \approx -30$, is somewhat too large, but it at least agrees in sign
with ours.

The fits reported above give also the coefficients $a_n(w) \equiv a_2(w) b_n$. 
In Ref.~\CITE{BN-97} it was claimed that $a_2(w)\approx 0$ for 
$w = 0.505838$. We can verify here this result. More importantly, 
we can test the renormalization-group prediction $a_n(w) = a_2(w) b_n$,
by verifying that not only does $a_2(w)$ approximately vanish, but that
the same property holds for the coefficient $a_3(w)$ [we are not precise 
enough to estimate reliably $a_4(w)$].
From the fits we obtain for $w = 0.505838$:
\begin{eqnarray}
  {a_2(w)\over A_2^*} &=& 0.08 \pm 0.02, \\
  {a_3(w)\over A_3^*} &=& 0.2  \pm 0.1 \; .
\end{eqnarray}
These results are not fully consistent with those of Ref.~\CITE{BN-97}.
Indeed, we find that, for $w = 0.505838$, $a_n(w)$ is not zero within error
bars.
Still, $w = 0.505838$ is very close to the optimal value $w^*$ for which
$a_2(w^*) = 0$.
Indeed, for $w = 0.505838$ corrections are a factor-of-10 smaller than those 
occurring for $w = 0.375$ and $w = 0.775$ and a factor-of-20 
smaller that those occuring in SAWs with $\beta = 0$ and $\beta = 0.1$,
the values used in Ref.~\CITE{PH-05}. Our best estimate for $w^*$ is
$w^* = 0.48\pm 0.02$. We have also recomputed the optimal $p$ 
introduced in Ref.~\CITE{PH-05}, which gives the optimal 
combination of SAW data corresponding to $\beta = 0$ and 
$\beta = 0.1$. We obtain $p_{\rm opt} = 0.52\pm 0.07$.

\begin{figure}
\centerline{\epsfig{file=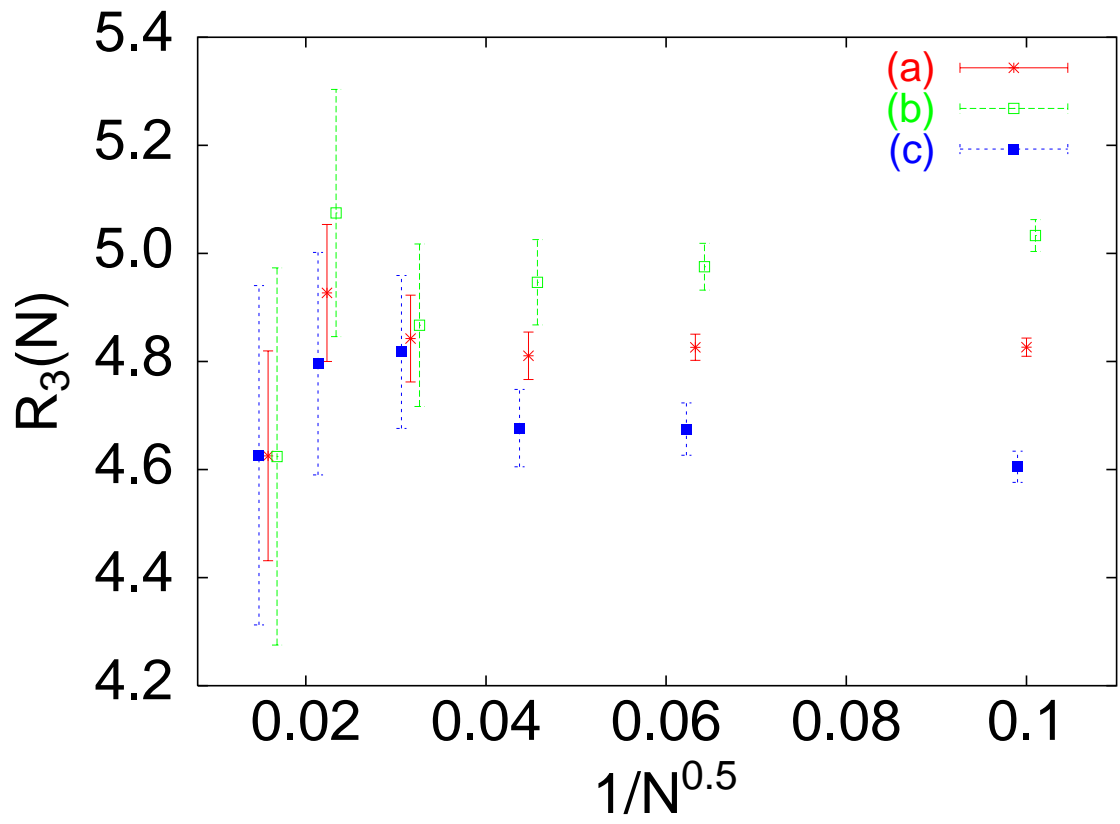,angle=-90,width=16truecm}}
\vspace{1cm}
\caption{Estimates of $R_3(N)$ vs $1/N^{0.5}$ for three different cases:
(a) $w_1 = 0.775$, $w_2 = 0.375$,
(b) $w_1 = 0.775$, $w_2 = 0.505838$;
(c) $w_1 = 0.375$, $w_2 = 0.505838$. }
\label{fig:b3}
\end{figure}

Finally, we compute the universal scaling-correction coefficient $b_3$.
We use the same method as described in Ref.~\CITE{PH-05}. We define
\begin{equation}
R_n(N) \equiv {A_n(N,w_1) - A_n(N,w_2)\over A_2(N,w_1) - A_2(N,w_2)},
\end{equation}
which should scale asymptotically as\cite{PH-05}
\begin{equation}
R_n(N) = b_n + {d_n(w_1,w_2)\over N^{\Delta_{\rm eff}}},
\end{equation}
with $\Delta_{\rm eff} = 0.5\pm 0.1$. We use the three possible 
choices of $w_1$ and $w_2$, verifying
the universality of the large-$N$ behavior of 
$R_3(N)$. In Fig.~\ref{fig:b3} we report $R_3(N)$ for the different cases.
It is clear that asymptotically all quantities 
converge to the same value, as predicted by the renormalization group. 
A fit of the data gives 
\begin{eqnarray}
   b_3 &=&  4.75\pm 0.30,
\end{eqnarray}
where the error includes the statistical error and the systematic error
due to the uncertainty on $\Delta_{\rm eff}$. 
In principle the same analysis can 
be applied to $A_4$, but here errors are so large that no reliable estimate
can be obtained.

\section{Osmotic pressure} \label{sec5}

The results of the previous Section allow us to determine the 
osmotic pressure in the dilute regime. Indeed, neglecting terms of 
order $c^4$ we can write 
\begin{equation}
Z = {\beta \Pi\over c} \approx 
     1 + 1.313 \Phi_p + 0.559 \Phi_p^2  - 0.122 \Phi_p^3 + 
     k_\Phi (\Phi_p + 1.13 \Phi_p^2 +
     {b}_{4,\Phi} \Phi_p^3 ), 
\label{Zris}
\end{equation}
where we have introduced the polymer packing fraction
\begin{equation}
\Phi_p \equiv {4 \pi R^3_g\over 3} c = {4 \pi R^3_g\over 3} {N_A\over M} \rho,
\end{equation}
$R^2_g$ is the zero-density radius of gyration, 
$N_A$ the Avogadro number, $M$ the molar mass of the polymer, 
$c$ and $\rho$ the number density and the ponderal concentration respectively.
Equivalently, we can write
\begin{equation}
Z \approx 1 + X  + 0.324 X^2 - 0.054 X^3 + k_X (X^2 + b_{4,X} X^3), 
\label{ZX}
\end{equation}
where $X \equiv B_2 c$, thereby avoiding any reference to the radius of 
gyration. The constants ${b}_{4,\Phi}$ and $b_{4,X}$ depend on $b_4$
for which we have only a rough estimate. If we trust the result
$b_4 = -2\pm 11$ obtained in Sec.~\ref{sec4}, we have 
${b}_{4,\Phi} = - 0.1 \pm 0.6$ and $b_{4,X} = 0.4 \pm 1.7$. 
The parameters $k_\Phi = 3 a_2/(4 \pi N^{\Delta})$ and 
$k_X \approx 0.0392 a_2/N^{\Delta}$ are nonuniversal and 
depend on the degree of polymerization. 

\begin{figure}
\centerline{\epsfig{file=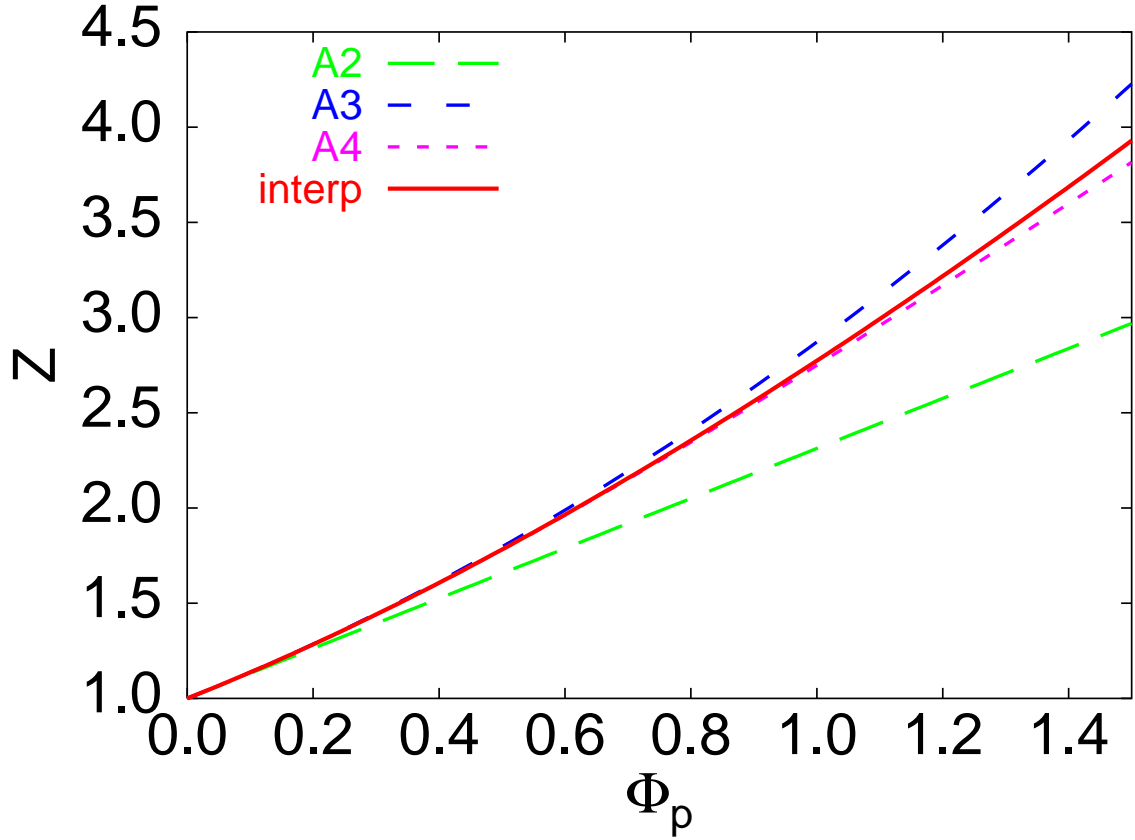,angle=-90,width=16truecm}}
\vspace{1cm}
\caption{Compressibility factor $Z$ vs $\Phi_p$. We report four different 
approximations: (A2) it corresponds to Eq.~(\ref{Zris}) truncated to order 
$\Phi_p$; (A3) truncation to order $\Phi_p^2$;
(A4) truncation to order $\Phi_p^3$;
(interp) interpolation formula (\ref{Zfinal}).
In all cases $k_\Phi = 0$. }
\label{fig:Zdilute}
\end{figure}

In Fig.~\ref{fig:Zdilute} 
we plot Eq.~(\ref{Zris}) for $k_\Phi = 0$ (scaling limit).
It is evident that $Z$ is linear in $\Phi_p$ only for very small 
values of $\Phi_p$, say $\Phi_p\lesssim 0.2$. For larger values of 
$\Phi_p$ inclusion of the higher-order terms is crucial. Note that 
approximations (A3) and (A4) in Fig.~\ref{fig:Zdilute} give very close
predictions for $Z$, indicating that in the region 
$\Phi_p\lesssim 1$ the virial expansion gives reasonably accurate results,
with errors of the order of a few percent.
Eqs.~\reff{Zris} and \reff{ZX} depend on a single nonuniversal parameter
that allows us to take into account the leading corrections to scaling 
due to the finite degree of polymerization. In practice, $k_\Phi$ and 
$k_X$ can be determined by requiring the measured value of $Z$ at a given
value of $\Phi_p$ or $X$ to agree with expression 
\reff{Zris} or \reff{ZX}. Then, all parameters are fixed and 
Eqs.~\reff{Zris} and \reff{ZX} predict $Z$ in the whole dilute regime 
$\Phi_p\lesssim 1$, $X \lesssim 1$.

To extend Eq.~\reff{Zris} or Eq.~\reff{ZX} 
into the semidilute regime, we must modify
them to take into account the asymptotic behavior 
in the scaling limit \cite{deGennes-79}
\begin{equation}
Z \sim \Phi_p^{1/(3\nu - 1)} \sim \Phi_p^{1.311} \sim X^{1.311}\; .
\label{Zlargec}
\end{equation}
Moreover, a proper resummation is necessary. Since the 
expansion of $Z^{3\nu - 1}=Z^{0.763}$ is alternating in sign, 
we will resum this quantity by using a Pad\'e approximant
that behaves as $\Phi_p$ for large concentrations.
We write therefore (for $k_\Phi = 0$) 
\begin{equation}
Z \approx \left( {1 + 1.5260 \Phi_p + 0.7954 \Phi_p^2 \over 
        1 + 0.5245 \Phi_p} 
     \right)^{1.311} .
\label{Zfinal}
\end{equation}
The numerical coefficients have been obtained by requiring 
this expression for $Z$ to reproduce Eq.~\reff{Zris} to order $\Phi_p^3$.
An analogous expression in terms of $X$ can be obtained by using 
$X \approx 1.313 \Phi_p$ in the scaling limit.

\begin{figure}
\centerline{\epsfig{file=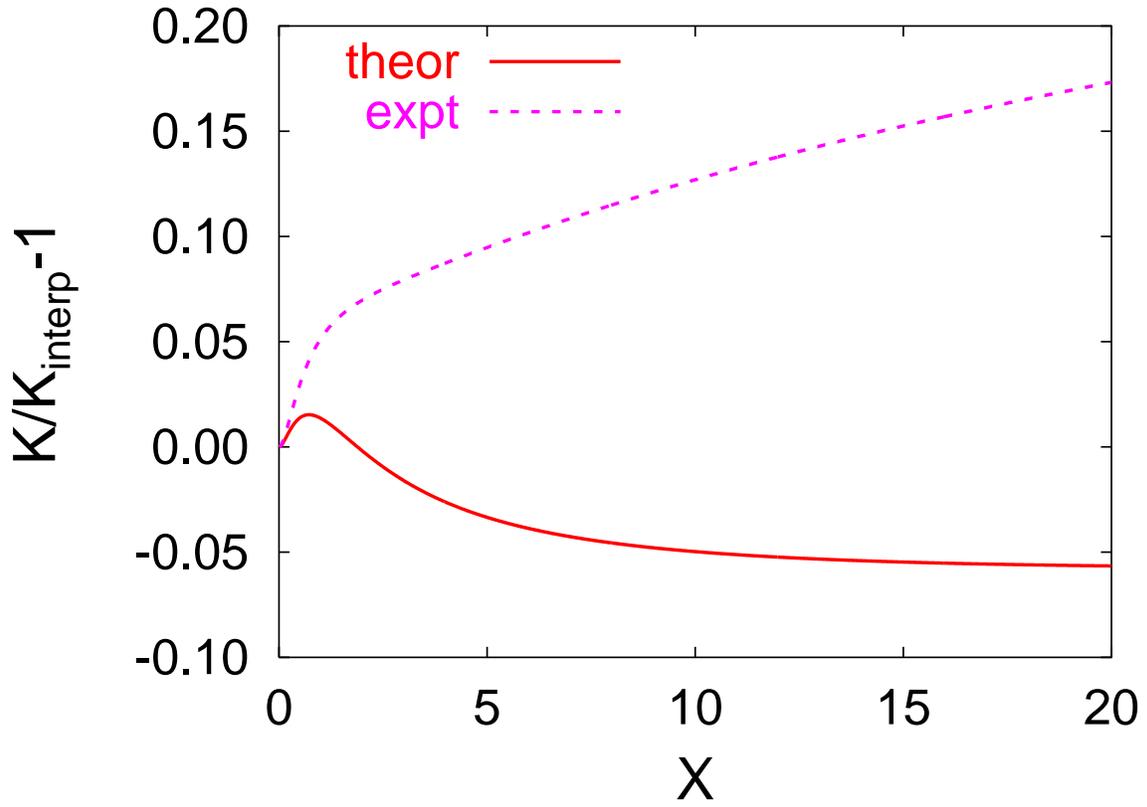,angle=-90,width=16truecm}}
\vspace{1cm}
\caption{Comparison of the compressibility $K$ obtained by using 
the interpolation formula \reff{Zfinal} ($K_{\rm interp}$) and 
two different approximations presented in Ref.~\protect\CITE{MBLFG-93}:
one is based on one-loop field theory (``theor"), one is obtained 
by fitting different sets of data for polystyrene (``expt").}
\label{fig:diff}
\end{figure}

Eq.~\reff{Zfinal} is of course very accurate in the dilute regime since it 
reproduces exactly the virial expansion \reff{Zris} 
(see Fig.~\ref{fig:Zdilute}). We must now assess the error for larger values of 
$\Phi_p$. We first compare with the renormalization-group predictions of 
Ref.~\CITE{MBLFG-93}. They give a simple parametrization of their 
one-loop $\epsilon$-expansion results for the 
compressibility $K\equiv (M/RT){\partial \Pi/\partial \rho}=
{\partial \beta\Pi/\partial c}$, which can be measured directly in scattering 
experiments (Eqs. 6 and 7 of Ref.~\CITE{MBLFG-93} with 
$p = 0.32$ and $q = 0.42$). In Fig.~\ref{fig:diff} we plot (``theor") the 
quantity $K_{\rm theor}/K_{\rm interp} - 1$, where $K_{\rm theor}$ is the 
prediction of Ref.~\CITE{MBLFG-93} and $K_{\rm interp}$ the expression
derived from Eq.~\reff{Zfinal}. The two predictions are close and,
for $X\to\infty$, the difference is approximately 5\%. We can
also compare with the field-theory results of Refs.~\CITE{dCN-82,Schaefer-99}.
For $\Phi_p\to\infty$, they predict $Z\approx 1.51 \Phi_p^{1.31}$
(Ref.~\CITE{dCN-82}) and $Z\approx 1.99 \Phi_p^{1.309}$ 
(Ref.~\CITE{Schaefer-99}, Sect. 17.4.1),
to be compared with $Z\approx 1.73 \Phi_p^{1.311}$ obtained by using 
Eq.~(\ref{Zfinal}). An interpolation formula\cite{footZinterp} 
for $Z$ is also given
in Ref.~\CITE{Schaefer-99}. It is in full agreement with ours up to 
$\Phi_p\approx 3$ (differences are less than 1\%). Then, the discrepancy 
increases, rising to 5\% for $\Phi_p\approx 10$.
These comparisons indicate that 
Eq.~\reff{Zfinal} gives a pressure that is slightly different (5-10\% 
at most) from the 
field-theory predictions. These differences should not be taken seriously,
since one-loop field-theory estimates have at most a 10-20\% precision,
as is clear from the results for $A_{3,22}$. We can also compare with the 
numerical results of  Ref.~\CITE{ALH-04}. 
Expression (\ref{Zfinal}) describes them reasonably well
(differences less than 10\%).

\begin{figure}
\centerline{\epsfig{file=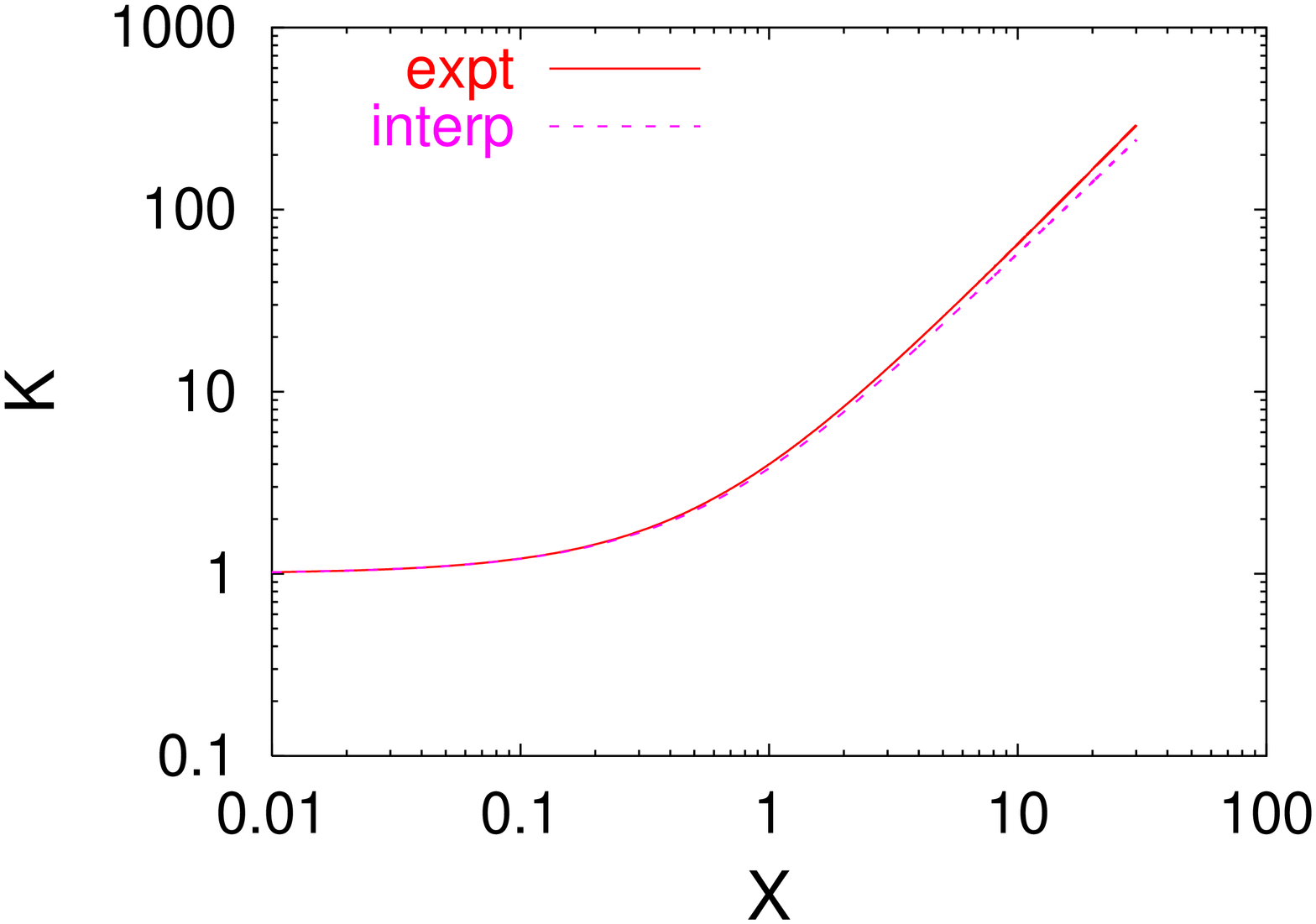,angle=-90,width=16truecm}}
\vspace{1cm}
\caption{Comparison of the compressibility $K$ obtained by using 
the interpolation formula \reff{Zfinal} (``interp") and 
by fitting different sets of data for polystyrene (``expt") 
(taken from Ref.~\protect\CITE{MBLFG-93}).
}
\label{fig:expt}
\end{figure}

We now compare our prediction with experiments. Ref.~\CITE{NKKN-81} quotes 
$Z\approx 1.50 \Phi_p^{1.32}$  for $\Phi_p\to \infty$, that 
apparently indicates that we are slightly overestimating the pressure.
Note however that the same data give 
$Z \approx 1 + 1.12 \Phi_p$ for $\Phi_p\to 0$ [compare with 
Eq.~\reff{Zris}], 
indicating that there are large scaling and/or polydispersity corrections.
Opposite conclusions are reached by comparing with compressibility results
for polystyrene.  Ref.~\CITE{MBLFG-93} gives
an empirical expression for the compressibility $K$ that fits well 
several sets of data for polystyrene
(Eqs. 6 and 7 of Ref.~\CITE{MBLFG-93} with 
$p = 0.39$ and $q = 0.46$). In Fig.~\ref{fig:expt} we report the experimental
results together with the prediction obtained by using Eq.~\reff{Zfinal}
(this figure is analogous to Fig. 1 of Ref.~\CITE{MBLFG-93}). 
Our expression follows quite closely the experimental data, though 
the experimental
compressibility is larger than our prediction, as can be better 
seen in Fig.~\ref{fig:diff} where we report (``expt")
$K_{\rm expt}/K_{\rm interp} - 1$.
Again, this discrepancy should not be taken too
seriously, since the experimental data do not satisfy the correct 
asymptotic behavior: they give $Z \sim X^{1.39}\sim \Phi_p^{1.39}$, to
be compared with the theoretical prediction \reff{Zlargec}. Thus, the 
discrepancy we observe could well be explained by scaling corrections and 
polydispersity effects.

Eq.~\reff{Zris} applies of course only to situations in which 
the solution is in the good-solvent regime. Close to the $\theta$
point, corrections are particularly strong and cannot be parametrized by a 
single coefficient $k_\Phi$. In this case, one can use the strategy 
proposed in Ref.~\CITE{PH-05}. Work in this direction is in progress.

\vskip 1truecm

The authors thank Tom Kennedy for providing his efficient simulation code 
for lattice self-avoiding walks.

\appendix

\section{Determination of the virial coefficients}

In order to evaluate the $n$-th order virial coefficient $B_{n}$ we need
to perform a summation over ${\mathbb Z}^{3(n-1)}$. For this purpose
we use the hit-or-miss algorithm discussed in Ref.~\CITE{LMS-95} for 
$B_2$. The algorithm can be trivially generalized to higher-order
virial coefficients.  We consider a walk 
$W$, with monomers ${\mathbf r}_0$,$\ldots$, ${\mathbf r}_N$,
starting at the origin (${\mathbf r}_0={\mathbf 0}$), and
define $\alpha_j^+(W)$ and $\alpha_j^-(W)$ as the maximum and 
minimum value of the $j$-th coordinate among the points of the walk.
Then, given two walks $W_1$ and $W_2$ we define
\begin{eqnarray}
S_{12} &\equiv & 
    [m_{1,12},M_{1,12}]  \times
    [m_{2,12},M_{2,12}]  \times
    [m_{3,12},M_{3,12}]  
\nonumber \\
    &=& 
      [\alpha_1^-(W_1) - \alpha_1^+(W_2),\alpha_1^+(W_1) - \alpha_1^-(W_2)] 
    \times
      [\alpha_2^-(W_1) - \alpha_2^+(W_2),\alpha_2^+(W_1) - \alpha_2^-(W_2)] 
\nonumber \\
    && 
    \times
      [\alpha_3^-(W_1) - \alpha_3^+(W_2),\alpha_3^+(W_1) - \alpha_3^-(W_2)] \; ,
\end{eqnarray}
and, given three walks $W_1$, $W_2$, and $W_3$, we define 
\begin{eqnarray}
S_{12,3} &\equiv &
   [m_{1,13} + m_{1,32},M_{1,13} + M_{1,32}]  \times
   [m_{2,13} + m_{2,32},M_{2,13} + M_{2,32}]  \nonumber \\
   && \times
   [m_{3,13} + m_{3,32},M_{3,13} + M_{3,32}]  \; .
\end{eqnarray}
It is easy to understand the rationale behind these definitions.
If walk $W_1$ starts in the origin and walk $W_2$ is translated and 
starts in a lattice point that does not belong to $S_{12}$,
then $W_1$ and the translated $W_2$ do not intersect each other, so that
the corresponding Mayer function $f_{12}$ vanishes.
Analogously, if  walk $W_1$ starts in the origin and walk $W_2$ is 
translated and starts in a lattice point that does not belong to 
$S_{12,3}$ there is no translation of $W_3$ such that 
the translated $W_3$ intersects both $W_1$ and $W_2$. This guarantees 
that in the calculation of the virial coefficient the product 
$f_{13} f_{23}$ always vanishes.
With these definitions the sums that need to be computed for 
$I_2$, $I_3$, and $I_4$ can be written as 
\begin{eqnarray}
I_2: && \sum_{{\mathbf r}_{12} \in S_{12}} f_{12}(\bz,{\mathbf r}_{12}),
\nonumber 
\\
I_3: && \sum_{{\mathbf r}_{12} \in S_{12}} 
        \sum_{{\mathbf r}_{13} \in S_{13}}
         f_{12}(\bz,{\mathbf r}_{12})
         f_{13}(\bz,{\mathbf r}_{13})
         f_{23}({\mathbf r}_{12},{\mathbf r}_{13}),
\nonumber 
\\
I_4: && \sum_{{\mathbf r}_{12} \in {D}_{12}}
        \sum_{{\mathbf r}_{13} \in {D}_{13}}
        \sum_{{\mathbf r}_{14} \in {D}_{14}}
\nonumber \\
   && \qquad \hphantom{+}
       [f_{12}(\bz,{\mathbf r}_{12}) f_{23}({\mathbf r}_{12},{\mathbf r}_{13})
         f_{34}({\mathbf r}_{13},{\mathbf r}_{14})
         f_{14}(\bz,{\mathbf r}_{14}) 
            (1 + 
              f_{13}(\bz,{\mathbf r}_{13}) + 
              f_{24}({\mathbf r}_{12},{\mathbf r}_{14}) ) 
\nonumber \\
   && \qquad + 
         f_{13}(\bz,{\mathbf r}_{13}) f_{14}(\bz,{\mathbf r}_{14}) 
         f_{23}({\mathbf r}_{12},{\mathbf r}_{13})
         f_{24}({\mathbf r}_{12},{\mathbf r}_{14}) 
            (1 + 
              f_{12}(\bz,{\mathbf r}_{12}) +
              f_{34}({\mathbf r}_{13},{\mathbf r}_{14})) 
\nonumber \\
   && \qquad + 
         f_{12}(\bz,{\mathbf r}_{12}) f_{13}(\bz,{\mathbf r}_{13})
         f_{24}({\mathbf r}_{12},{\mathbf r}_{14})
         f_{34}({\mathbf r}_{13},{\mathbf r}_{14}) 
             (1 + 
               f_{14}(\bz,{\mathbf r}_{14}) + 
               f_{23}({\mathbf r}_{12},{\mathbf r}_{13}) ) 
\nonumber \\
    && \qquad + 
         f_{12}(\bz,{\mathbf r}_{12}) f_{13}(\bz,{\mathbf r}_{13})
         f_{14}(\bz,{\mathbf r}_{14}) f_{23}({\mathbf r}_{12},{\mathbf r}_{13})
         f_{24}({\mathbf r}_{12},{\mathbf r}_{14})
         f_{34}({\mathbf r}_{13},{\mathbf r}_{14}) ],
\label{Virial-compute}
\end{eqnarray}
with 
\begin{eqnarray}
{D}_{12} &=& {S}_{12,3} \cap {S}_{12,4},
\nonumber\\
{D}_{13} &=& {S}_{13,2} \cap {S}_{13,4},
\nonumber\\
{D}_{14} &=& {S}_{14,2} \cap {S}_{14,3}.
\label{choise}
\end{eqnarray}
Here $f_{ij}({\mathbf r},{\mathbf s})$ is the Mayer function 
computed for walk $i$ starting in ${\mathbf r}$ and 
walk $j$ starting in ${\mathbf s}$. 

Eq.~\reff{Virial-compute} shows that the computation of the 
virial coefficients requires the calculation of finite sums. They
can be determined by a simple hit-or-miss procedure that provides 
an unbiased estimate. For instance,
in order to compute the contribution to $I_2$ 
we extract randomly $\ell$  vectors ${\mathbf r}^{(a)}_{12}
\in {S}_{12}$ ($a=1,\cdots,\ell$), and compute
\begin{eqnarray}
&& { V(S_{12})\over \ell}
\sum_{a=1}^\ell f_{12}({\mathbf 0},{\mathbf r}_{12}^{(a)}),
\nonumber\\
&& V(S_{12}) = (M_{1,12}-m_{1,12}+1)(M_{2,12}-m_{2,12}+1)
(M_{3,12}-m_{3,12}+1).
\end{eqnarray}
These considerations easily generalize to higher-order coefficients.

The other contributions $T_1$, $T_2$, $T_3$, and $T_4$ do not require any 
additional work, since they factorize in products of independent sums.
For instance, to determine $T_1$ we need to compute 
\begin{equation}
\sum_{{\mathbf r}_{12},{\mathbf r}_{13}}
  f_{12}(\bz,{\mathbf r}_{12}) 
  f_{13}(\bz,{\mathbf r}_{13})  = 
  \left[ \sum_{{\mathbf r}_{12}\in S_{12}} f_{12}(\bz,{\mathbf r}_{12})\right]
  \left[ \sum_{{\mathbf r}_{13}\in S_{13}} f_{13}(\bz,{\mathbf r}_{13})\right]
\end{equation}
The two sums are independent and can be evaluated separately as we did
for the contribution to $I_2$. 

In the calculation of the $n$-th virial coefficient with 
the hit-or-miss method 
we need to choose a point in a $3(n-1)$-dimensional lattice parallelopiped. 
This is done by using $3(n-1)$ random numbers. For the fourth virial 
coefficient, 9 random numbers are needed to compute each contribution.
If the random number generator one is using has nonnegligible short-range
correlations (this is the case of congruential generators,
see Ref.~\CITE{Knuth}), 
the results may be incorrect. Therefore, we took particular 
care in the choice of the random number generator. 
We considered four different random number generators: 
a congruential generator with prime modulus
\[
    y_n = {\rm mod}\,(16807 y_{n-1},2^{31} - 1);
\]
a 48-bit congruential generator
\[
    z_n = {\rm mod}\,(31167285 z_{n-1} + 1,2^{48});
\]
a 32-bit shift-register generator with very long period
\[
    t_n = t_{n-1029}\, {\tt XOR}\, t_{n-2281},
\]
where {\tt XOR} is the exclusive-or bitwise operation;
the 32-bit Parisi-Rapuano generator\cite{PR-85}
\[
    x_n = {\rm mod}\,(x_{n-24} + x_{n-55},2^{32}) \qquad\qquad
    r_n = x_n \, {\tt XOR}\, x_{n-61}.
\]
In order to compute the virial coefficients we must choose
one or more lattice points in a given three-dimensional parallelopiped. 
For this purpose we generate
four uniform random numbers $a$, $a_1$, $a_2$, and $a_3$ in [0,1):
\begin{eqnarray}
a &=& {\rm mod}\,(t_n + 2 y_n,2^{32}) \times 2^{-32}, \nonumber \\
a_1 &=& {\rm mod}\,(r_n + 2 y_{n+1},2^{32}) \times 2^{-32}, \nonumber \\
a_2 &=& {\rm mod}\,(r_{n+1} + 2 y_{n+2},2^{32}) \times 2^{-32}, \nonumber \\
a_3 &=& z_n\times 2^{-48}.
\end{eqnarray}
Number $a$ is used to determine a random permutation $\sigma$ of three 
elements. Then, we consider ${\mathbf v} \equiv (v_1,v_2,v_3) = 
(a_{\sigma(1)}, a_{\sigma(2)}, a_{\sigma(3)})$. 
A random lattice point in $[m_1,M_1]\times [m_2,M_2]\times [m_3,M_3]$
is  $(m_1 + \lfloor v_1  (M_1 - m_1 + 1)\rfloor,
      m_2 + \lfloor v_2  (M_2 - m_2 + 1)\rfloor,
      m_3 + \lfloor v_3  (M_3 - m_3 + 1)\rfloor)$.
As a check, we computed the virial coefficients for hard spheres using the 
hit-or-miss method. We obtain:
\begin{eqnarray}
B_2/V &=& 0.499993\pm 0.000069, \\
B_3/V^2 &=& 0.156231\pm 0.000062, \\
B_4/V^3 &=& 0.035785\pm 0.000072 ,
\end{eqnarray}
where $V$ is the volume of the sphere. These estimates should be compared 
with the exact results\cite{HMD-76}
 0.5, 0.15625, 0.035869$\ldots$ They are in perfect
agreement. Thus, we are confident that our final results, that are much less
precise than those reported above, are not affected by any bias due to the 
random number generator.

\begin{table}
\caption{Estimates of the ratio $A_2$ for different values of $N$ and $w$.}
\label{tableA2}
\begin{tabular}{cccc}
 & $w=0.375$ & $w=0.505838$ &  $w=0.775$ \\
\hline
$N=100$ & $5.20511(83)$ & $5.49049(58)$ & $5.79591(85)$ \\
$N=250$ & $5.31755(81)$ & $5.50395(63)$ & $5.69318(64)$ \\
$N=500$ & $5.37200(93)$ & $5.50609(68)$ & $5.63798(93)$ \\
$N=1000$& $5.4116(14)$  & $5.5048(10)$  & $5.5988(14)$ \\
$N=2000$& $5.4365(14)$  & $5.50610(93)$ & $5.5676(14)$ \\
$N=4000$& $5.4545(13)$  & $5.5027(10)$  & $5.5466(14)$ \\
$N=8000$& $5.4682(14)$  & $5.5051(10)$  & $5.5324(15)$ \\
\end{tabular}
\end{table}

\begin{table}
\caption{Estimates of the ratio $A_3$ for different values of $N$ and $w$.}
\label{tableA3}
\begin{tabular}{cccc}
 & $w=0.375$ & $w=0.505838$ &  $w=0.775$ \\
\hline
$N=100$ & $8.4794(54)$ & $9.7936(41)$ & $11.3308(61)$ \\
$N=250$ & $8.9534(59)$ & $9.8248(49)$ & $10.7663(48)$ \\
$N=500$ & $9.2024(65)$ & $9.8295(46)$ & $10.4819(74)$ \\
$N=1000$& $9.3724(74)$ & $9.8214(71)$ & $10.2789(89)$ \\
$N=2000$& $9.4760(99)$ & $9.8098(65)$ & $10.1219(91)$ \\
$N=4000$& $9.595(11)$  & $9.8180(71)$ & $10.021(11)$ \\
$N=8000$& $9.647(10)$  & $9.8062(74)$  & $9.945(11)$ \\
\end{tabular}
\end{table}

\begin{table}
\caption{Estimates of the ratio $A_4$ for different values of $N$ and $w$.}
\label{tableA4}
\begin{tabular}{cccc}
 & $w=0.375$ & $w=0.505838$ &  $w=0.775$ \\
\hline
$N=100$ & $-8.64(33)$ & $-8.64(30)$ & $-7.82(53)$ \\
$N=250$ & $-8.86(49)$ & $-8.61(34)$ & $-7.91(41)$ \\
$N=500$ & $-7.84(48)$ & $-9.14(37)$ & $-8.55(64)$ \\
$N=1000$& $-9.24(86)$ & $-9.23(65)$ & $-7.92(87)$ \\
$N=2000$& $-8.30(94)$ & $-8.96(69)$ & $-8.5(1.0)$ \\
$N=4000$& $-8.8(1.0)$ & $-8.19(70)$ & $-9.6(1.0)$ \\
$N=8000$& $-8.0(1.0)$ & $-9.92(77)$ & $-10.1(1.1)$ \\
\end{tabular}
\end{table}

\begin{thebibliography}{199}
\bibitem{footexpt}
In experimental works the virial coefficients are usually defined 
from the expansion of $Z/M$ in terms of the ponderal concentration $\rho$,
$Z/M = 1/M + \sum_n B_{n+1}^{\rm expt} \rho^n$,
where $M$ is the molar mass of the polymer. The coefficients 
$B_{n}^{\rm expt}$ are related to those
we have defined by $B_{n}^{\rm expt} = N_A^{n-1} B_n M^{-n}$, where 
$N_A$ is the Avogadro number.
\bibitem{footA4}
In Sec.~17.3.2 of Ref.~\CITE{Schaefer-99} field theory is used 
to obtain a prediction for the virial expansion. In terms of 
the variable $\hat{s} = 4 \pi c R_g^3/(3\times 1.169)$, 
$Z = 1 + 1.433 \hat{s} + 1.065 \hat{s}^2 - 0.648 \hat{s}^3 + \cdots$.
This implies $Z = 1 + 5.13 c R_g^3 + 13.7 (c R_g^3)^2 - 
30 (c R_g^3)^3 + \cdots$. Note that also $A_3^*$ is considerably
different from our result.
\bibitem{footZinterp}
Eq.~17.52 of Ref.~\CITE{Schaefer-99} gives the following expression
for $Z$: $Z = 1 + 1.314 \Phi_p (H_1/H_2)^{0.309}$, 
$H_1 = 1 + 2.15 \Phi_p + 1.00 \Phi_p^2$,
$H_2 = 1 + 0.51 \Phi_p$. The variable $\hat{s}$ used in 
Ref.~\CITE{Schaefer-99} is related to 
$\Phi_p$ by $\Phi_p = 1.169 \hat{s}$ (see Sec. 13.3.2). For $\Phi_p\to \infty$
it predicts $Z \approx 1.61 \Phi_p^{1.309}$, that is slightly lower
than the prediction $Z \approx 1.99 \Phi_p^{1.309}$, reported in 
Sec.~17.4.1 of the same reference, indicating that the expected error
on the parametrization should be of the order of 10\%.
\end{thebibliography}
\end{document}